\documentclass{Interspeech}

\interspeechcameraready

\usepackage{cite}
\usepackage{amsmath,amssymb,amsfonts}
\usepackage{algorithmic}
\usepackage{graphicx}
\usepackage{makecell}
\usepackage{textcomp}
\usepackage[table]{xcolor}

\usepackage{url}
\usepackage{bm}
\usepackage{booktabs}
\usepackage{multirow,multicol}
\usepackage{enumitem}
\usepackage{etoolbox}
\usepackage{hyperref}
\usepackage{bibspacing}  
\usepackage{comment}
\begin{document}

\newcommand{\MethodName}{LSCodec}

\title{\MethodName: Low-Bitrate and Speaker-Decoupled Discrete Speech Codec}

\author[affiliation={1,2}]{Yiwei}{Guo}
\author[affiliation={1,2}]{Zhihan}{Li}
\author[affiliation={1,2}]{Chenpeng}{Du}
\author[affiliation={1,2}]{Hankun}{Wang}
\author[affiliation={1,2}]{Xie}{Chen}
\author[affiliation={1,2}]{Kai}{Yu}


\affiliation{X-LANCE Lab, MoE Kay Lab of Artificial Intelligence, School of Computer Science}{Shanghai Jiao Tong University}{China}
\affiliation{}{Jiangsu Key Lab of Language Computing}{China}
\email{yiwei.guo@sjtu.edu.cn, kai.yu@sjtu.edu.cn}
\keywords{Discrete speech tokens, speech codecs, low-bitrate coding, speaker decoupling, speech language modeling}

\newcommand{\blue}[1]{\textcolor{blue}{#1}}

\maketitle


\begin{abstract}
Although discrete speech tokens have exhibited strong potential for language model-based speech generation, their high bitrates and redundant timbre information restrict the development of such models. In this work, we propose \MethodName, a discrete speech codec that has both \underline{l}ow bitrate and \underline{s}peaker decoupling ability. \MethodName~adopts a multi-stage unsupervised training framework with a speaker perturbation technique. A continuous information bottleneck is first established, followed by vector quantization that produces a discrete speaker-decoupled space. A discrete token vocoder finally refines acoustic details from \MethodName. By reconstruction evaluations, \MethodName~demonstrates superior intelligibility and audio quality with only a \textbf{single codebook} and smaller vocabulary size than baselines. 
Voice conversion and speaker probing experiments prove the excellent speaker disentanglement of \MethodName, and ablation study verifies the effectiveness of the proposed training framework.

\end{abstract}

\section{Introduction}
{\let\thefootnote\relax\footnotetext{\vspace{-0.1in}Kai Yu is the corresponding author.}}

Currently, discrete speech tokens have laid a foundation for speech generation tasks~\cite{VQTTS,valle,yang2024uniaudio} since they can be treated in a similar manner with the prevailing language models (LMs).
By their objectives, discrete speech tokens can be divided into acoustic tokens and semantic tokens~\cite{yang2024towards}.
Acoustic tokens, or \textit{speech codecs} like EnCodec~\cite{encodec} and DAC~\cite{kumar2024high}, aim to reconstruct audio perfectly with a vector quantization (VQ) module.
While these tokens achieve remarkable audio decompression quality, they usually require multiple VQ layers which leads to extra complex designs in speech LMs, such as coarse-to-fine modeling~\cite{valle}, delay pattern~\cite{voicecraft} or nested prediction~\cite{yang2024uniaudio}.
Tokens with lower bitrates are favored not only for transmission, but also for simplifying the speech LM architectures to reduce cost.

Different from other aspects of information in speech, speaker timbre is a global trait nearly orthogonal to contents and prosody.
In most acoustic tokens, timbre is repeatedly encoded across timesteps.
We believe this leads to a considerable waste of coding efficiency.
Contrarily, semantic tokens are derived from discriminative tasks, such as self-supervised learning (SSL)~\cite{vq-wav2vec,hsu2021hubert,baevski2020wav2vec,chen2022wavlm}, that focus more on contents other than acoustics.
Thus these tokens usually have fewer codebooks than acoustic ones, resulting in lower bitrate.
However, semantic tokens may discard much paralinguistic information including both time-invariant timbre and time-variant prosody~\cite{mousavi2024dasb,shi24h_interspeech}.
As prosody is an important part of emotion and semantics, this makes semantic tokens not ideal for speech generation neither.

Therefore, it is highly valuable to develop a low-bitrate and speaker-disentangled discrete speech token.
The bitrate of speech tokens are determined by three factors: number of quantizers $Q$, frame rate $F$ and vocabulary size $V$ for each quantizer\footnote{\vspace{-0.15in}Bitrates are computed by $Q\times F\times \lceil\log_2V\rceil$ bps.}.
With a smaller $F$, the length mismatch of speech and text modalities can be mitigated.
With a smaller $Q$ and $V$, speech LMs will have simpler training objectives.
A small $V$ can also make more utilization of acoustic BPEs~\cite{shen2024acoustic,guo2025recent}.
Less speaker variations also reduces the modeling difficulty of speech LMs to concentrate only on time-variant features, and speaker timbre can be rendered after LM generation.

Fortunately, lowering compression bitrate and removing speaker information benefit each other mutually.
It is common to involve designs that reduce global information in speech tokens.
In \cite{jiang2023disentangled,ticodec,singlecodec,zheng2024freecodec}, a speaker encoder or global-level extractor is applied to provide timbre-like features into the decoder.
The natural information bottleneck in the VQ layer helps reduce time-invariant features in the encoded tokens.
Particularly, \cite{ticodec,singlecodec,zheng2024freecodec} explore the scenario of only one VQ codebook.
However, no explicit decoupling technique is introduced, which does not guarantee the effectiveness of the bottleneck.
Single-Codec also sacrifices stream-ability due to its non-causal encoder.
Explicit speaker decoupling is explored via gradient reversal with or without supervised data~\cite{SSVC,facodec}, contrastive loss~\cite{qian2022contentvec} or swapped prediction~\cite{spin}, but these works do not pursue low-bitrate coding.
Several other single-codebook codecs have been developed~\cite{ji2024wavtokenizer,parker2025scaling,wu2024ts3} with decent reconstruction performance and low bitrates, but without speaker decoupling.

In this work, we propose \MethodName, the first effort to explicitly consider speaker decoupling that forms a single-codebook low-bitrate speech token unsupervisedly.
\MethodName~adopts a three-stage learning process that firstly trains a speech variational autoencoder (VAE) with speaker perturbation to achieve reasonable speaker disentanglement and temporal compression.
Then a VQ module is added to form a VQ-VAE where \MethodName~tokens are produced.
Finally, a discrete token vocoder is individually trained on the \MethodName~tokens to refine audio quality.
We use a simple stretching-based perturbation method to introduce explicit removal of timbre.
We train two versions of \MethodName~under 50 and 25Hz frame rates, using only a single codebook and a small vocabulary size for 24kHz speech.
For \MethodName-50Hz, the vocabulary size is $V=300$ with a resulting bitrate of 0.45kbps, while for \MethodName-25Hz we choose $V=1024$ with bitrate 0.25kbps.
We show by experiments that \MethodName~owns superior reconstruction performance albeit having a single codebook and lower bitrate than the baselines.
Voice conversion (VC) and speaker probing experiments further verify that \MethodName~achieves remarkable speaker disentanglement.
Hence it paves the way for future speech LMs due to its low bitrate, high compactness and speaker invariance.



Audio demos are available online\footnote{{https://cantabile-kwok.github.io/LSCodec/}}.

\section{\MethodName: {Low-Bitrate Speaker-Decoupled Codec}}

\begin{figure}
    \centering
    \includegraphics[width=0.98\linewidth]{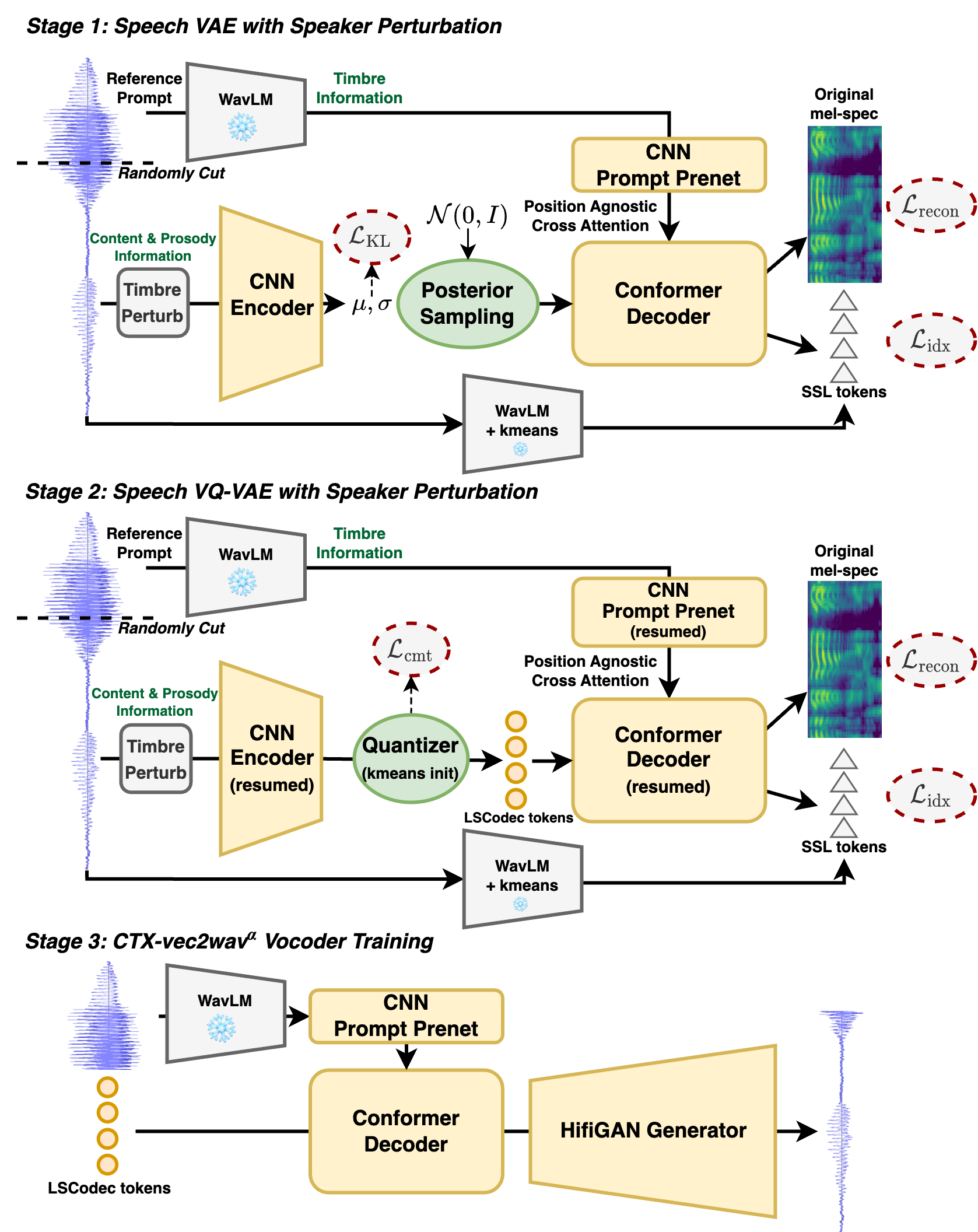}
    \caption{Architecture and Training Diagram of \MethodName.}
    \label{fig:main}
    \vspace{-0.2in}
\end{figure}

The architecture and training diagram of \MethodName~are illustrated in Fig.\ref{fig:main}.
\MethodName~adopts a three-stage learning process where speaker perturbation is applied on the first two stages. 
We elaborate on the perturbation method and three training stages in this section.

\vspace{-0.07in}
\subsection{Speaker Perturbation}
\vspace{-0.03in}
\label{sec:perturb}

To explicitly disentangle discrete speech tokens and speaker information, 
appropriate modification on speaker timbre is necessary before tokenization, while keeping content and prosodic variations retained.
Several techniques have been explored in previous study, such as formant \& F0 scaling, random equalizers~\cite{qian2022contentvec}, and vocal tract length perturbation~\cite{chan2022speechsplit2}.
In this work, we use a simple time stretching approach that produces time-aligned perturbed speech with only pitch and timbre changes.
Given a coefficient $\beta$, our perturbation process starts with a rate-based speed-up effect to scale the total duration of an utterance by $\beta$ times.
This operation changes pitch and formant positions, hereby altering the timbre.
Then, a pitch-preserving tempo effect is applied to re-stretch the utterance to its original duration, via the WSOLA algorithm~\cite{verhelst1993overlap}.
This perturbation\footnote{The perturbation process can be efficiently implemented via SoX.} only changes the global pitch position and timbre features while retaining the content and pitch variations.
Therefore, to reconstruct speech, timbre must be learned from additional inputs instead of the perturbed segments.
With a strong information bottleneck in the reconstruction process, speaker can be explicitly disentangled from the speech tokens.
In training stage, $\beta$ is independently sampled within an interval centered at 1 for every utterance to provide randomized variations.

\vspace{-0.07in}
\subsection{Training Stage 1: Speech VAE}
\vspace{-0.03in}
The first training stage of \MethodName~is a speech VAE in a continuous space.
Given an utterance waveform, we randomly cut it into a reference prompt and content segment.
The content segment $\bm x$ is passed to the perturbation algorithm in Sec.\ref{sec:perturb} with a random $\beta$.
Then, a convolutional neural network (CNN) encoder compresses the signal in time domain and outputs isotropic Gaussian posteriors $\mathcal N(\bm \mu_t,\bm\sigma_t^2)$ for each frame $t$.
The frame rate is controlled by the CNN strides.
The sampled posteriors are fed to a Conformer~\cite{conformer}-based decoder.
To form an information bottleneck before the decoder like \cite{qian2019autovc,chan2022speechsplit2}, we provide the decoder with sufficient timbre information.
We use the position-agnostic cross attention~\cite{du2024unicats,li2024sef} to feed timbre from prompt segments into the decoder, as it exhibits superior timbre controllability.
To provide more discriminative timbre features, we use the WavLM~\cite{chen2022wavlm} SSL model to extract hidden embeddings from reference prompt, inspired by its widely-verified advantage on speaker verification~\cite{jung2024espnet,superb}.
As the reference and content segments are different and the cross attention mechanism is position-agnostic, the reference prompt can almost only provide timbre information.
With sufficient timbre provided in the decoder and a perturbed input, the information bottleneck formed by VAE naturally disentangles the posterior and timbre to a degree. 
The WavLM embeddings are fed to a CNN prompt prenet before entering the decoder.

The decoder predicts two features: mel-spectrogram and SSL semantic tokens from the original un-perturbed content segment.
In other words, the VAE model performs multi-task learning that simultaneously regresses towards acoustic mel features and classifies towards SSL semantic tokens in a fixed vocabulary size.
The ground truth SSL tokens are extracted using WavLM~\cite{chen2022wavlm} and k-means clustering.
The prediction of mel-spectrograms ensures reconstruction ability of the bottleneck features.
As semantic tokens retain rich content information with a compact discrete space, the SSL token prediction task is crucial for guiding the bottleneck features to encode sufficient contents.
Meanwhile, prosody information is largely damaged in the discrete WavLM tokens~\cite{mousavi2024dasb}, so keeping the mel prediction task ensures prosody to be encoded in the bottleneck.
Note that SSL models do not require labels to train, so the whole training pipeline is still unsupervised unlike FACodec~\cite{facodec}.

Therefore in this stage, the training loss contains a KL loss $\mathcal L_{\text{KL}}=\frac 1T \sum_{t=1}^T D_{\text{KL}}(\mathcal N(\bm \mu_t,\bm\sigma^2_t)||\mathcal N(\bm 0, \bm 1))$, an L1 loss of predicted mel features $\mathcal L_{\text{recon}}$ and cross entropy loss of predicted SSL indexes $\mathcal L_{\text{idx}}$.
Three losses are weighted with $\gamma_{\text{KL}},\gamma_{\text{recon}}$ and $\gamma_{\text{idx}}$, respectively.
Due to the information bottleneck in this VAE, speaker timbre can already be partially removed in a continuous space.

\vspace{-0.07in}
\subsection{Training Stage 2: Speech VQ-VAE}
\vspace{-0.03in}
Next, we inject a VQ layer with codebook size $V$ into the trained VAE to obtain the desired \MethodName~tokens.
We first extract the Gaussian means $\bm \mu$ from the VAE using a portion of training data, and perform a $V$-centroid k-means clustering for initializing the codebook.
The architecture of VQ-VAE remains almost identical to the VAE in stage 1, and we resume the encoder, decoder and prompt prenet parameters before training.
The VQ layer quantizes the encoder outputs to the codebook entry with the smallest Euclidean distance in $V$ candidates.
Straight-through estimator~\cite{van2017neural} is applied for gradient back-propagation.
The quantized tokens are mapped to code-vectors before entering the decoder.
The codebook in the VQ layer is updated via exponential moving average (EMA), following \cite{encodec}.
The training criterion also inherits that from stage 1, except for replacing the KL loss with the commitment loss~\cite{van2017neural} $\mathcal L_{\text{cmt}}$ in the VQ layer.
Note that the dimensions for previous Gaussian variances are disabled in this stage.
The loss $\mathcal L_{\text{cmt}}$ is weighted via $\gamma_{\text{cmt}}$.

With a VQ layer and resumed parameters, this VQ-VAE constructs a discrete space based on the insufficiently speaker-decoupled continuous space from stage 1.
The information bottleneck thus further restricts timbre from being encoded.
The indexes after the VQ layer are the desired LSCodec tokens.

\vspace{-0.04in}
\subsection{Training Stage 3: Vocoder CTX-vec2wav$^\alpha$}
As synthesizing waveforms from predicted mel-spectrograms often yields degraded audio quality, we additionally train a specialized vocoder from the obtained discrete tokens like Vocos~\cite{siuzdak2024vocos}.
Since timbre is decently removed in the \MethodName~tokens, we resort to CTX-vec2wav~\cite{du2024unicats} that exhibits strong timbre controllability~\cite{li2024sef}.
To be consistent with the previous two stages, we develop an improved version CTX-vec2wav$^\alpha$ that receives the same WavLM outputs as timbre features instead of mel-spectrograms from reference prompts.
Other parts of the vocoder follow the original CTX-vec2wav.
With this vocoder, acoustic details can be refined to produce high-fidelity waveforms.
This vocoder is trained from scratch.
The final model can also serve as a high-quality VC model, where timbre can be well altered by providing different reference prompts.

\vspace{-0.1in}
\section{Experimental Results}

\subsection{Data and Model Setup}
We use LibriTTS~\cite{libritts} for model training, which is an English corpus with 585 hours of 24kHz speech data.
To cover enough timbre variations, we use all of the train splits that contain about 2500 speakers.
We discard all utterances shorter than 6s, remaining around 360 hours of training data.
The prompts are randomly cut with lengths sampled between one third and one half of the original duration.

We train two versions of \MethodName~in 50Hz and 25Hz frame rates.
The CNN encoders contain 11 and 12 residual blocks respectively and 512 hidden dimensions.
In stage 1, the output of encoder is 128 dimensional before being evenly split and projected into $\bm\mu,\bm\sigma^2$. 
In stage 2, only the first half dimensions of encoder output are used for quantizing and decoding.
The code vectors is then 64-dimensional.
Codebook is updated with EMA weight $\gamma=0.99$.
Code expiration is enabled only after 5000 update steps.
The Conformer decoder contains 2 Conformer blocks each with 2 heads and 184 attention dimensions.
The prompt prenet has four CNN blocks with scaled residual connections, where the hidden dimensions are 128, 256 and 512 before being fed to cross attentions.

We use the output of the sixth layer from a pretrained WavLM-Large model\footnote{\scriptsize{https://github.com/microsoft/unilm/tree/master/wavlm}} to provide timbre features.
In the SSL token prediction task, we use the same pretrained WavLM and perform 2048-centroid k-means clustering on around a 83-hour subset of our training data.
As this SSL feature resides in 50Hz, in the 25Hz version of \MethodName, we repeat each token by 2 times along the temporal axis before the Conformer decoder.
The 80-bin cepstral-normalized mel-spectrograms for reconstruction also have a frame rate of 50Hz.

Each stage of \MethodName~is trained up to 200 epochs, with loss weights $\gamma_{\text{recon}}=\gamma_{\text{KL}}=60$, $\gamma_{\text{idx}}=2$, $\gamma_{\text{cmt}}=1$. 
The perturbation coefficient $\beta$ is sampled uniformly in $[0.8,1.2]$.
Other hyper-parameters and the training of CTX-vec2wav$^\alpha$ follow that in \cite{du2024unicats}.

\vspace{-0.08in}
\subsection{Speech Reconstruction}
\vspace{-0.02in}

\begin{table}[]
\centering
\caption{Reconstruction results. Bitrates are presented in kbps. Parentheses denote model versions, and ``-'' means not comparable because of sampling rate.}
\vspace{-0.07in}
\label{tab:recon}
\resizebox{\columnwidth}{!}{
\begin{tabular}{@{}lccccc@{}}
\toprule
\textbf{Discrete Speech Token} & \textbf{Bitrate$\downarrow$} & \textbf{WER$\downarrow$} & \textbf{GPE$\downarrow$} & \textbf{SECS$\uparrow$}& \textbf{MOS$\uparrow$} \\ \midrule
Ground truth & - & 1.20 & 0.00 & 1.000 & 4.71$\pm$0.06 \\
\midrule
\multicolumn{6}{l}{\textit{Semantic (SSL) tokens with CTX-vec2wav$^\alpha$ vocoder}} \\ 
\textbf{wav2vec2.0} (Large) & 0.83 & 3.24 & 2.92& 0.947 & 4.45$\pm$0.08 \\
\textbf{WavLM} (Large) & 0.55 & 1.67 & 17.94 & 0.934 & 4.36$\pm$0.07 \\ \midrule
\multicolumn{5}{l}{\textit{Acoustic tokens with low bitrates at 16kHz}} \\
\textbf{SemantiCodec} & 0.63 & 4.16 & 2.15 & 0.926 & - \\
\textbf{LLM-Codec} & 0.85 & 6.25 & 1.86 & 0.919 & - \\
\textbf{Stable-Codec} (2 posthoc) & 0.70 & 5.06 & \textbf{1.73} & 0.889 & - \\
\multicolumn{6}{l}{\textit{Acoustic tokens with low bitrates at 24kHz}} \\ 
\textbf{DAC} (First VQ only) & 0.75 & 25.76 & 8.20 & 0.746 & unintelligible \\
\textbf{TiCodec} (1VQ) & 0.75 & 10.44 & 2.24 & 0.857 &  3.56$\pm$0.06 \\
\textbf{WavTokenizer} (Small, 40Hz) & 0.48 & 7.86 & {1.89} & 0.907 & 4.14$\pm$0.07 \\
\rowcolor{gray!20} 
\textbf{\MethodName}~(50Hz, $V$=300) & 0.45 & \textbf{3.33} & 2.42 & \textbf{0.954} & \textbf{4.49$\pm$0.07} \\
\rowcolor{gray!20} 
\textbf{\MethodName}~(25Hz, $V$=1024) & \textbf{0.25} & 5.46 & 2.80 & 0.945 & 4.41$\pm$0.07 \\
\bottomrule
\end{tabular}
}
\vspace{-0.13in}
\end{table}

To evaluate the reconstruction ability of \MethodName~against other low-bitrate speech tokens, we use the testset-B in \cite{du2024unicats} that contains 500 utterances from unseen speakers in LibriTTS test-clean split.
Since low-bitrate speech coding cannot fully restore every detail in the original waveform, we emphasize on comparing the preservation of key information: content, prosody, and speaker timbre.
We use word error rate (WER, \%) to measure intelligibility and gross pitch error (GPE, \%) to measure prosody preservation
WERs are computed using NeMo-ASR\footnote{\scriptsize{https://huggingface.co/nvidia/stt\_en\_fastConformer\_transducer\_large}} with the ground truth transcriptions. 
GPE stands for the ratio of pitch values larger than 20\% relative error from the ground truth.
We also compute speaker similarity cosine distance (SECS) from Resemblyzer\footnote{\scriptsize{https://github.com/resemble-ai/Resemblyzer}} between reconstructed and original utterances.
We conduct MOS tests as a subjective evaluation of audio quality.

The baselines include semantic SSL tokens with CTX-vec2wav$^\alpha$ vocoders, where wav2vec 2.0~\cite{baevski2020wav2vec} tokens are obtained from its inner quantizer, and WavLM tokens are 2048-centroid cluster indexes from the last layer\footnote{Same as that used in SSL token prediction task.}.
To enable a fair comparison among the low-bitrate acoustic speech codecs, we compare those with bitrate lower than 1kbps.
The baselines then include DAC~\cite{kumar2024high} (first VQ only),  TiCodec~\cite{ticodec}, WavTokenizer-small~\cite{ji2024wavtokenizer} as single-codebook baselines, and SemantiCodec~\cite{semanticodec}, LLM-Codec~\cite{yang2024uniaudio15}, Stable-Codec~\cite{parker2025scaling} as multi-codebook baselines.
Official checkpoints are used for these baselines.
We also report the bitrate of these baselines to indicate model versions.
For SemantiCodec, we use the 50Hz token rate version with semantic vocabulary size 4096.
For Stable-Codec, we use 2 posthoc bottlenecks.
MOS tests are conducted among tokens at 24kHz sampling rate and decent reconstruction quality.

The results are presented in Table \ref{tab:recon}, which reveal that \MethodName-50Hz~exhibits significantly better WER, SECS and subjective audio quality than other acoustic token baselines although having even a lower bitrate.
Although \MethodName~has a slightly higher GPE than TiCodec and WavTokenizer, the improvement on WER is still worthwhile, since a 1\% difference in GPE is hardly noticeable in human perception, but WER matters more in downstream tasks.
Compared to semantic tokens, \MethodName-50Hz has comparable WER with wav2vec 2.0 tokens but better GPE and SECS, and both \MethodName~versions have a much better GPE than HuBERT tokens.
Note that wav2vec 2.0 has much worse speaker decoupling than \MethodName~as will be shown in later experiments.
The 25Hz version of \MethodName~has a minor degradation than the 50Hz version, since its bitrate is only 55\% of the latter.
\MethodName-25Hz still has competitive WER and SECS among baselines.
These findings suggest that \MethodName~owns remarkable reconstruction performance using a very low bitrate.

\vspace{-0.06in}
\subsection{Any-to-Any Voice Conversion}
\vspace{-0.02in}

\begin{table}[]
\centering
\caption{VC results. Bitrates are presented in kbps.}
\vspace{-0.1in}
\label{tab:vc}
\resizebox{\columnwidth}{!}{
\begin{tabular}{@{}lcccc@{}}
\toprule
\textbf{Discrete Speech Token} & \textbf{Bitrate$\downarrow$} & \textbf{WER$\downarrow$} & \textbf{SECS$\uparrow$} & \textbf{P.Corr$\uparrow$} \\ \midrule
\multicolumn{5}{l}{\textit{Semantic (SSL) tokens with CTX-vec2wav$^\alpha$ vocoder}} \\ 
\textbf{wav2vec2.0} (50Hz) & 0.83 & 4.40 & 0.814 & 0.759 \\
\textbf{WavLM} (50Hz) & 0.55 & 1.92 & 0.872 & 0.374 \\ \midrule
\multicolumn{5}{l}{\textit{Acoustic tokens with VC ability}} \\ 
\textbf{FACodec} (80Hz, w/ \textit{detail}) & 4.80 & \textbf{1.57} & 0.774 & 0.583 \\
\textbf{FACodec} (80Hz, w/o \textit{detail}) & 1.60 & {2.01} & 0.822 & 0.558 \\
\textbf{TiCodec} (75Hz, 1VQ) & 0.75 & 9.40 & 0.714 & 0.785 \\
\textbf{TiCodec} (75Hz, 2VQ) & 1.50 & {2.62} & 0.642 & 0.886 \\
\rowcolor{gray!20} 
\textbf{\MethodName}~(50Hz, $V$=300) & 0.45 & 4.04 & 0.852 & 0.697 \\
\rowcolor{gray!20} 
\textbf{\MethodName}~(25Hz, $V$=1024) & \textbf{0.25} & 6.32 & \textbf{0.853} & 0.753 \\ \bottomrule
\end{tabular}
}
\vspace{-0.2in}
\end{table}

While reconstruction evaluations reflect the preservation of content and prosody information, speaker disentanglement can be evaluated by VC.
In this section, we randomly assign a different target speaker for every source utterance in testset-B for VC task.
We measure SECS between the target and converted utterances.
Higher SECS means better timbre controllability of the tokens.
In addition, we also compute the correlation coefficient of pitch contours (P.Corr) between the source and converted speech.
This allows us to have a better picture of a VC model both from timbre and prosody similarity.
Both SECS and P.Corr range from -1 to 1.
Note that P.Corr becomes meaningless when SECS is low.
We include the same SSL tokens with CTX-vec2wav$^\alpha$ for comparison as in Table \ref{tab:recon}.
As only few codec models have the ability of VC, we include TiCodec (both 1 VQ and 4 VQ variants) and FACodec~\cite{facodec} to comparison. 
The \textit{detail} tokens in FACodec can either be discarded or preserved, which will lead to different VC outcomes.

The results are shown in Table \ref{tab:vc}.
Concretely, \MethodName~has better speaker similarity in VC than other acoustic tokens, especially FACodec with supervised decoupling of speakers.
We find that TiCodec has unsatisfactory performance on speaker similarity, which leads to inordinately high P.Corr since converted speech is very similar to the source one.
This demonstrates the difficulty of disentangling speakers only via implicit bottleneck methods.
In VC scenarios, \MethodName~still exhibits lower WERs than TiCodec.
Compared with semantic tokens, \MethodName~owns notably higher SECS than wav2vec 2.0 tokens and P.Corr than HuBERT, achieving a better balance between the two metrics.
These findings reflect that speaker timbre in \MethodName~tokens have been removed well with little harm on prosody, which results in competitive VC ability.

\subsection{Speaker Information Probing}
\begin{figure}
    \centering
    \includegraphics[width=0.99\linewidth]{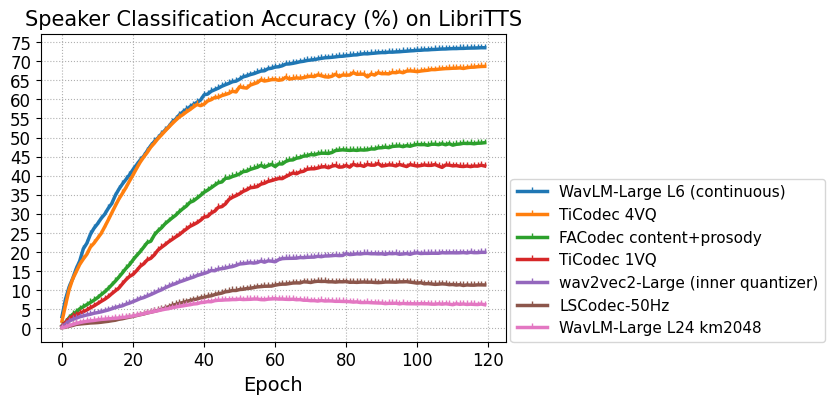}
    \vspace{-0.15in}
    \caption{Accuracy curve in speaker probing experiments.}
    \label{fig:probe}
\end{figure}

We also conduct speaker probing experiments as a more straight proof of the disentanglement ability.
We do so by speaker classification on LibriTTS with the Xvector~\cite{xv} network as the speaker classifier.
We use code-vectors associated with each token to feed the classifier.
We leave out 10\% utterances from the LibriTTS training set for validation.
Lower classification accuracy means less speaker information present in the quantized space.
The accuracy curves on the validation set along with training epochs are visualized in Fig.\ref{fig:probe}.
Among acoustic tokens in Table.\ref{tab:vc}, \MethodName~achieves the lowest probing accuracy, demonstrating its success on speaker timbre removal.
For semantic tokens, \MethodName~still outperforms wav2vec2.0 tokens.
Since \MethodName~retains prosody information, it is reasonable that it gets higher accuracy than WavLM tokens.

\vspace{-0.05in}
\subsection{Ablation Study}
\vspace{-0.04in}

We conduct another ablation study to verify the effectiveness of speaker perturbation, SSL token prediction task and the multi-stage training framework.
We only include the first two training stages where perturbation and SSL token prediction take place.
The results are shown in Table \ref{tab:ablation}. 
Compared with the \MethodName-50Hz with $\beta\in[0.8,1.2]$, canceling or restricting speaker perturbation leads to worse WER and SECS in both stages.
This shows that the model is sacrificing local contents for more global information to decrease the loss values.
Similar patterns can be found when $\beta$ ranges are enlarged, since this brings too much damage to data.
Canceling the SSL token prediction loss has a significant impact on WERs in both stages.
Directly training from the VQ-VAE stage also has a degraded performance in both metrics.
Hence, we conclude that the proposed training framework effectively produces the desired properties of \MethodName.



\begin{table}[]
\centering
\caption{Ablation study on training techniques}
\vspace{-0.1in}
\label{tab:ablation}
\resizebox{\columnwidth}{!}{
\begin{tabular}{@{}llcccc@{}}
\toprule
 \multirow{2}{*}{\textbf{Ablation}} & \multirow{2}{*}{\textbf{Configuration}} & \multicolumn{2}{c}{\textbf{S1 (VAE)}} & \multicolumn{2}{c}{\textbf{S2 (VQ-VAE)}} \\
\cmidrule(l){3-4}\cmidrule(l){5-6} & & \textbf{WER$\downarrow$} & \textbf{SECS$\uparrow$} & \textbf{WER$\downarrow$} & \textbf{SECS$\uparrow$} \\
 \midrule
 Original \MethodName-50Hz & With $\mathcal L_{\text{idx}}$, $\beta\in[0.8,1.2]$ & 4.96 & 0.811 & 3.39 & 0.817 \\
\midrule
\multirow{3}{*}{On perturbation} & No Perturb & 5.48 & 0.799 & 3.97 & 0.806 \\
& $\beta\in[0.9,1.1]$ & 4.78 & 0.804 & 3.78 & 0.807 \\
& $\beta\in[0.7,1.3]$ & 5.40 & 0.809 & 3.62 & 0.809 \\ 
\midrule
On SSL token prediction & No $\mathcal L_{\text{idx}}$ & 11.22 & 0.811 & 8.93 & 0.816 \\
\midrule
On multi-stage training & No stage 1 training & - & - & 3.84 & 0.800 \\
\bottomrule
\end{tabular}
}
\vspace{-0.2in}
\end{table}

\vspace{-0.05in}
\section{Conclusion}
\vspace{-0.03in}
We propose \MethodName, a low-bitrate and speaker-decoupled speech codec with a highly compact representation space.
\MethodName~constructs a multi-stage training framework with speaker perturbation.
A single-codebook VQ layer is applied after a VAE that disentangles speaker in a continuous space.
A token vocoder is then trained upon the quantized codes.
Experiments prove that \MethodName~achieves strong reconstruction and VC ability given a bitrate as low as 0.45 and 0.25kbps.
Probing experiments and ablation studies validate the proposed methods.
Meanwhile, stronger perturbation methods, better content preservation, and scaling ability can be explored as future work.

\newpage



\section{Acknowledgements}
This work is supported by China NSFC Project (No. 92370206), Shanghai Municipal Science and Technology Major Project (2021SHZDZX0102), and Key Research and Development Program of Jiangsu Province, China (No.BE2022059).


\bibliographystyle{IEEEtran}
\bibliography{refs}

\end{document}